\documentclass[aps,prl,twocolumn,showpacs,superscriptaddress,amsmath,amssymb]{revtex4}

\usepackage{graphicx}        
\usepackage{color,colordvi}
\usepackage{epstopdf}
\usepackage{dcolumn}
\usepackage{multirow}
\usepackage{color,colordvi}
\usepackage{bm}              
\usepackage{psfrag}
\usepackage{subfigure}
\usepackage{amsmath}
\usepackage{ulem,colordvi,amsmath,epsfig,float,color,subfigure}
\usepackage{textcomp}
\begin{document}
\title{Magnetism: the Driving Force of Order in CoPt. A First-Principles Study.}
\author{S. Karoui}
\affiliation{Laboratoire d'Etude des Microstructures, ONERA-CNRS, BP 72, 92322 Ch\^atillon Cedex, France}
\author{H. Amara}
\affiliation{Laboratoire d'Etude des Microstructures, ONERA-CNRS, BP 72, 92322 Ch\^atillon Cedex, France}
\author{B. Legrand}
\affiliation{Service de Recherches de M\'etallurgie Physique, DMN-SRMP, CEA Saclay, 91191 Gif-sur-Yvette, France}
\author{F. Ducastelle}
\affiliation{Laboratoire d'Etude des Microstructures, ONERA-CNRS, BP 72, 92322 Ch\^atillon Cedex, France}
\date{\today}

\begin{abstract}

CoPt or FePt equiatomic alloys order according to the tetragonal L1$_0$ structure which favors their strong magnetic anisotropy. Conversely magnetism can influence chemical ordering. We present here {\it ab initio} calculations of the stability of the L1$_0$ and L1$_2$ structures of Co-Pt alloys in their paramagnetic and ferromagnetic states. They show that magnetism strongly reinforces the ordering tendencies in this system. A simple tight-binding analysis allows us to account for this behavior in terms of some pertinent parameters.

\end{abstract}


\maketitle

Magnetism and chemical ordering are frequently coupled in alloys. On one hand, a strong magnetocrystalline anisotropy characteristic of equiatomic binary alloys like FePt or CoPt  is known to be due to an L1$_0$ order that alternates pure planes along the [001] direction~\cite{Coffey95,Alloyeau09, Blanc11, Tournus08, Andreazza10}. On the other hand, magnetism can influence phase stability and chemical ordering~\cite{Cadeville87,Bieber91}. Typical examples include the stability of the bcc $\alpha$ phase of iron~\cite{Pettifor95} and the phase diagram of FeCo~\cite{Abrikosov96,Rahaman11}. In the case of NiFe alloys involving neighboring late transition elements, magnetism plays an important role in the stability of ordered phases in a rather notable fashion~\cite{Ducastelle91, Ducastelle92}. It is suspected that this should also be true for FePt and CoPt alloys and, in the case of CoPt,  diffuse scattering as well as nuclear magnetic resonance experiments have established a clear relationship between magnetism and short range order \cite{Kentzinger00}. 

Furthermore, these alloys, in the form of nanometer-sized grains, are ideal candidates for high density magnetic storage applications (provided that the L1$_{0}$ ordered state is preserved). The systematic modeling of their thermodynamic properties, of the critical ordering temperature in particular \cite{Alloyeau09}, necessitates the development of multi-scale methods involving effective interatomic potentials. These potentials should include all the relevant physics of the alloy, while remaining simple enough to allow simulation of real life situations with hundreds or even thousands of atoms ~\cite{Rose81, Finnis84, Hardouin09, Rosato89, Mehl96, Jarvi09}. It then becomes imperative to determine whether magnetism governs energetic properties of bulk alloys and nanoalloys. Should this prove to be true, all interatomic potentials would need to include a magnetic term. Clearly, this is not the case of the potentials in current use  \cite{Ersen08, Muller07, Yang06, Rossi08, Qin10}.

The purpose of this Letter is to quantify the role of magnetism in the formation energies of Co-Pt alloys through non-magnetic and magnetic \textit{ab initio} electronic structure calculations; thus providing tangible grounds for a magnetic interatomic potential. 
In the bulk form, three ordered phases are known to exist in the Co-Pt system at  low temperature, corresponding to the stoichiometric concentrations of Co$_{3}$Pt, CoPt, and CoPt$_{3}$~\cite{Leroux88}. The ordered phase of CoPt is of the tetragonal L1$_0$ type. Co$_{3}$Pt and CoPt$_{3}$ are both L1$_{2}$ ordered phases of cubic symmetry. The bulk order-disorder transition of CoPt is equal to 1098 K \cite{Bouar03}. 

We carried out spin-polarized calculations in the framework of the Density Functional Theory (DFT) using the ABINIT code~\cite{Gonze02} with the Generalized Gradient Approximation (GGA) exchange correlations functionals. Core and valence electrons were represented by  a plane wave basis and the projector augmented wave (PAW) potentials~\cite{Dewaele08}. The adopted valence electronic configurations for Co and Pt are $3d^{8}4s^{1}$ and $5d^{9}6s^{1}$ respectively. All plane waves with energies below the cut-off energy were included in the basis set. The cut-off energies (16 eV, 22 eV, 22 eV for Co, Pt and CoPt, respectively) were chosen 25\% larger than the largest default cut-off of the element-specific potentials. Integrations over the Brillouin zone are based on a $ 20 \times 20\times 20$ Monkhorst-Pack 2D grid which is sufficiently fine to ensure the numerical convergence of all the calculated properties. All the structures were fully relaxed using the Broyden-Fletcher-Goldfarb-Shannon minimization. The cold smearing method was used for the Brillouin zone integration with a smearing parameter of 7$\times$10$^{-2}$ Ha leading to formation energies converged to within 10$^{-3}$ eV/at.  The calculations were performed at zero pressure; the relaxation of the atoms and the shape of the simulation cell are considered using the conjugate gradient minimization scheme. The atomic positions are relaxed until the forces on the atoms are reduced to within $10^{-7}$ Ha/Bohr.

\begin{figure}[htbp!]   
\begin{center}
\includegraphics[width=0.95\linewidth]{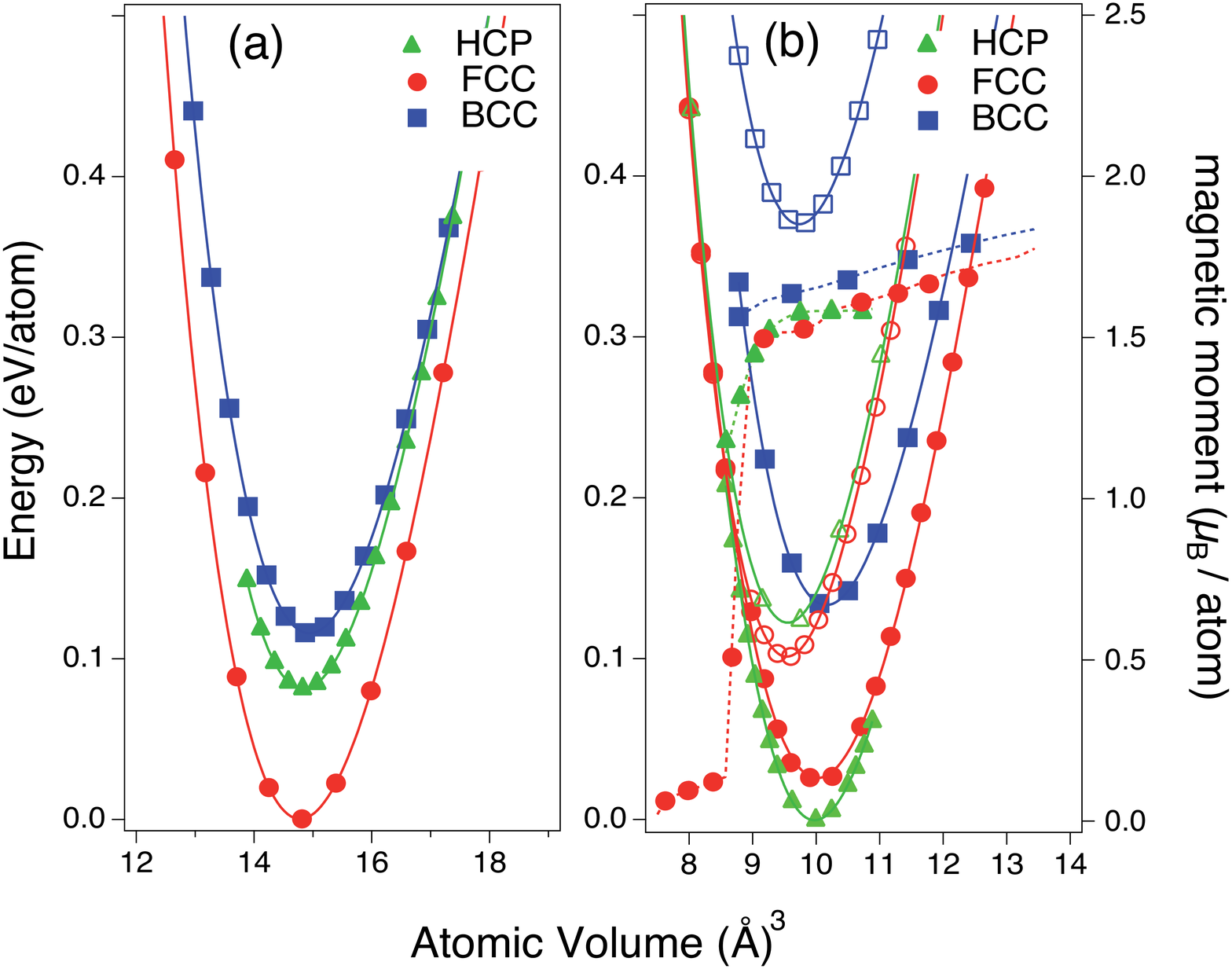}
\end{center}
\caption{Total energy as a function of atomic volume for (a) Pt and (b) Co as predicted by GGA calculations (open markers : non magnetic, full markers : ferromagnetic). For Co, the variation of the magnetic moment is also displayed.}
\label{Figure1}
\end{figure}

We first verified that face centered cubic (fcc), body centered cubic (bcc) and hexagonal close packed (hcp) structures were correctly reproduced in elemental Co and Pt. The relative stability of these various phases, as well as the influence of magnetism on the system can be determined from energy versus atomic volume curves plotted in Fig.\ \ref{Figure1}a for Pt and Fig.\ \ref{Figure1}b for Co. Concerning Co, the non-spin polarized calculations predict a fcc ground state while spin polarized calculations correct this  and reproduce the experimentally stable hcp phase (see Fig.\ \ref{Figure1}b). Such behaviors of non magnetic calculations have been highlighted in previous calculations~\cite{Cerny03}. Regarding the magnetic moment, as expected, it increases when the lattice is expanded and vanishes when it is reduced. The total energies of the ferromagnetically ordered bcc and fcc structures are slightly higher than the hcp ones, 0.12 eV/at and 0.01 eV/at excess energies per atom, respectively. Since the energy difference between fcc and hcp is relatively small and since the CoPt alloys are cubic or tetragonal according to the phase diagram, cobalt will be considered to be in its fcc phase in what follows.
 \begin{table}[htbp]
\caption{Physical properties of Co and Pt in the FCC structure (GGA calculations). Lattice parameter (a$_{0}$) and magnetic moment per atom ($m_{tot}$). DFT and experimental (when they exist) values are presented in brackets.}
\begin{ruledtabular}
\begin{tabular}{lcc}
& a$_{0}$ (\AA)  &$m_{tot}$  ($\mu_{B}$)\\ \hline

Co                          &  3.52 (3.53~\cite{Pearson})    &1.64 (1.62~\cite{Modak06})\\ 
Co$_{3}$Pt                  &    3.66 (3.66~\cite{Pearson})       &1.43 (1.45~\cite{Sipr08})  \\
CoPt (L1$_{0}$)                  & a=3.81 (3.81~\cite{Pearson})         &  1.14 (1.20~\cite{Cadeville87})\\
                                                &c/a=0.976 (0.973~\cite{Pearson})  & \\
CoPt (L1$_{1}$)                  & a=3.70  (3.80~\cite{Dannenberg09})&     1.08  \\
                                           &b/a=1.016 (1.017~\cite{Dannenberg09})&  \\
CoPt (A$_{2}$B$_{2}$)        &     3.71     &1.12  \\
CoPt$_{3}$                 &  3.86 (3.83~\cite{Pearson})       &0.73 (0.67~\cite{Cadeville87})  \\
Pt                &   3.97 (3.92\cite{Kittel})    & 0 \\
\end{tabular}
\end{ruledtabular}
\label{Data_CoPt}
\end{table}

As shown in Table ~\ref{Data_CoPt}, our results for physical and magnetic properties for the fcc structure are in agreement with both previous \textit{ab initio}, and when relevant, experimental data. We have calculated a magnetic moment within the LDA and GGA approximation equal to 1.54 $\mu_B$/at and 1.64 $\mu_B$/at, respectively. The LDA underestimation of the magnetic moment is clearly connected to the well-known underestimation of the lattice parameters generally obtained with LDA. Conversely, the usual overestimation of the lattice parameter by the GGA is not observed in the case of Co, leading to a good agreement between the experimental and the GGA determination of the magnetic moment. This leads us to prefer the use of GGA to model the CoPt system, even though Pt structural properties are slightly better reproduced by LDA. Regardless of the approximation, note that all structures for Pt are found to be non magnetic (NM), the ground-state structure being NM fcc as shown in Fig.~\ref{Figure1}a.

We now focus on ordered L1$_{0}$ and L1$_{2}$ alloys  which are known to exist in the CoPt system. The L1$_{0}$ phase exhibits uniaxial anisotropy with lattice parameters $a$ and $c$ about 3.81 and 3.71 \AA \ respectively. As was the case for elemental Co and Pt, the physical properties obtained for both alloyed structures are in excellent agreement with experimental data and other \textit{ab initio} results, as shown in Table~\ref{Data_CoPt}. Moreover, we have also considered a simplified geometry where  $c=a$; the magnetic moment was found to differ by 2 \% and the total energy by 0.02 eV/at. This suggests that the precise value of the $c/a$ ratio is irrelevant when the foci are the general physical properties of the structure. In particular,  its magnetocrystalline anisotropy is principally governed by the (anisotropic) chemical ordering \cite{Plumer01} more than by the deviation from unity of this ratio. Therefore, in the following we assume  that $c = a$.

\begin{figure}[htbp!]   
\begin{center}
\includegraphics[width=0.95\linewidth]{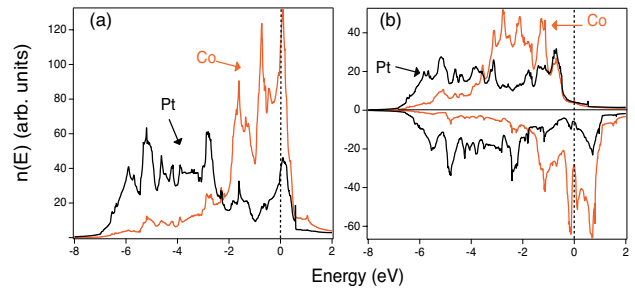}
\end{center}
\caption{Atom-projected electronic densities of states of CoPt L1$_{0}$: NM (a) and FM (b). }
\label{Figure2}
\end{figure}

In Fig.~\ref{Figure2}, we show the atom-projected electronic densities of states (PDOS) for the L1$_{0}$-non magnetic (NM) and L1$_{0}$-ferromagnetic (FM) phases of CoPt. As expected, the Pt band is wider than the Co band, about 8 eV and 5 eV respectively. The bands above $-7$ eV are predominantly of $d$ character and also show a large hybridization, in this case between the Co $3d$ and Pt $5d$ states.  In the paramagnetic phase, the Fermi level falls within a  Van Hove peak, which, according to the Stoner criterion, suggests a possible instability towards a ferromagnetic state.
Such is the case, in Fig.~\ref{Figure2}b, where a quasi-rigid shift between the majority (spin-up) and the minority (spin down) bands is observed. This alloy is a strong ferromagnet with a full majority spin band as was the case for elementary fcc or hcp Co. About 1 eV above the Fermi level, we observe a hybridization between Co and Pt $d$ states, leading to a small induced local magnetism on Pt, about 0.3 $\mu_B$/Pt atom. The PDOS for the L1$_{2}$ phase shows roughly the same behaviour. These calculations are consistent with previous works ~\cite{Koote91,Kashyap99,McLaren05}.

Similar studies were conducted in the alloyed phase at varying concentrations of Pt (Co$_{1-x}$Pt$_{x}$). The ordered phases L1$_{2}$ (Co$_{3}$Pt or CoPt$_{3}$) have been considered, as well as other equiatomic phases :  L1$_{1}$, and  the A$_{2}$B$_{2}$ phase \cite{Ducastelle91}. Our results are presented in Table~\ref{Data_CoPt}. For the L1$_{1}$ structure in particular, the $b/a$ ratio has been found equal to 1.016, closely comparable to the value 1.017 of previous  \textit{ab initio} calculations~\cite{Dannenberg09}. The total magnetic moment expressed per atom as in table I decreases as a function of the Pt concentration: this is due to the fact that the magnetic moment of Pt atoms is negligible, whereas the magnetic moment of Co atoms is almost independent of the Pt concentration, and even increases slightly with it, as already noticed \cite{Sipr08}.

Finally the enthalpies of formation ($\Delta H$, at $T=0 K$) of these alloys were determined:
\begin{equation}
\label{equation_dH}
\Delta H=\left[E_{\text{tot}}^{\text{Alloy}} (n\text{Co}, m\text{Pt})-nE_{\text{fcc}}^{\text{Co}}-mE_{\text{fcc}}^{\text{Pt}}\right]/(n+m)
\end{equation}
where $E_{\text{tot}}^{\text{Alloy}} (n\text{Co}, m\text{Pt})$ is the total energy of the mixed Co $+$ Pt system containing $n$ Co atoms and $m$ Pt atoms, and $E^{X}$ represents the energy per atom of the elemental form X (X= Co or Pt) in the appropriate reference state. 
\begin{figure}[htbp!]   
\begin{center}
\includegraphics[width=1.0\linewidth]{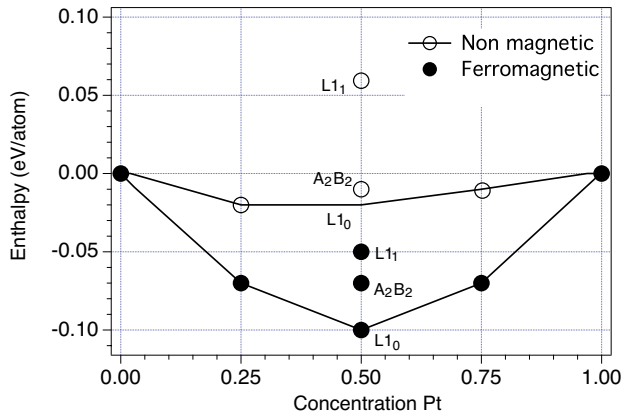}
\end{center}
\caption{Enthalpy of formation at 0K as predicted by GGA calculations : non magnetic and ferromagnetic. }
\label{Figure3}
\end{figure}
Thus, for the non spin polarized calculations, the reference states for the two elements are the non magnetic structures whereas for the spin polarized calculations we consider the ferromagnetic state for Co and the non magnetic state for Pt.

Fig.\ \ref{Figure3}  illustrates the importance of magnetism in assessing the stability of the ordered phases of the Co-Pt system. In non spin polarized calculations, $\Delta H$ is found to be very weak for any concentration. This result is incompatible with experimental observations of ordered structures having critical temperatures above 1000 K.  In the spin polarized calculations, $\Delta H$ takes negative values between -0.07 and -0.10 eV/at for the  L1$_{2}$ and L$1_{0}$ phases. These phases are therefore clearly stabilized by magnetic effects. The  L1$_{1}$ and $A_{2}B_{2}$ equiatomic structures  are also clearly stabilized by magnetic effects but are less stable than the L1$_0$ phase.  The effect of magnetism is more pronounced for the L1$_{1}$ phase than for the A$_{2}$B$_{2}$ and L1$_{0}$ phases, but not sufficient to modify the relative stability of the phases in comparison with the non spin polarized calculations. The energy differences between these three phases show that effective pair interactions beyond first neighbors are not negligible  \cite{Ducastelle91} and that our \textit{ab initio} calculations are successful in predicting L1$_0$ to be the most stable phase.

Our results for the L1$_0$ phase agree also with those obtained by Alam \textit{et al.}\ \cite{Alam10}; these authors (who did not considered other equiatomic structures) found a formation enthalpy about -0.09 eV/at for the ferromagnetic state, to be compared to our value about -0.10 eV/at  and the experimental value about -0.13 eV/at \cite{Hultgren73}. Their value for the paramagnetic state is slightly smaller than ours: -0.03 eV/at instead of $\simeq$ -0.02 eV/at. It should be mentioned that the uncertainties due the approximations made (LDA or GGA, pseudopotentials, KKR-CPA in the case of disordered states) are probably much larger. Recent calculations \cite{Chepulskii11} also agree on the order of magnitude of the formation enthalpy (about -0.1 eV/at for L1$_0$ ferromagnetic CoPt) but the role of magnetism is not discussed.

\begin{figure}[htbp!]   
\begin{center}
\includegraphics[width=0.95\linewidth]{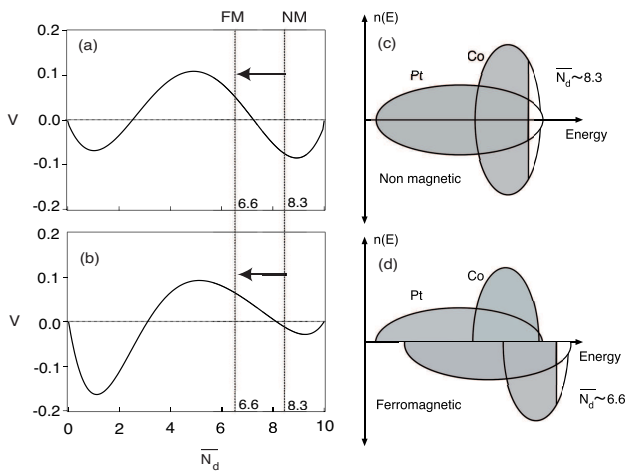}
\end{center}
\caption{Typical variation of $V$ as a function of the band-filling $N_{d}$ for an equiatomic composition. V $>0$ means an ordering tendency. (a)  $\delta_{d}$= 2 eV, $\delta_{nd}$ = 0 eV and (b) $ \delta_{d}$= 2 eV, $\delta_{nd}$ = 3 eV. Schematic representations of band filling for CoPt (c) non magnetic and (d) ferromagnetic. }
\label{Figure4}
\end{figure}

Even though \textit{ab initio} calculations are efficient in determining a quantitative description of an alloy with relatively high precision, general trends are best understood using simple models such as those provided by the tight-binding approximation \cite{Ducastelle91, Ducastelle92}. In the simplest model, the main alloy parameter is the diagonal disorder parameter, \textit{i.e.},  the difference in atomic $d$-levels $\delta_{d} = \epsilon_{d}^{B}- \epsilon_{d}^{A}$ compared to an average bandwidth. In this context, the main outcome of tight-binding studies is the justification of an Ising Hamiltonian for order phenomena in transition metals~\cite{Ducastelle76}. In addition, effective pair interaction between first neighbors ($V$) which dominates this Hamiltonian can be calculated through the Coherent Potential Approximation (CPA) for disordered alloys~\cite{Ducastelle91}. 

Fig.\ \ref{Figure4}.a  illustrates a typical variation of $V$ as a function of the average number of $d$-electrons $\overline{N}_{d}$ present in the system for an equiatomic alloy. Although for a real alloy $\overline{N}_{d}$ and the atomic concentration are related through $\overline{N}_{d}=cN^{A}_{d}+ (1-c)N^{B}_{d}$ for an $A_{c}B_{1-c}$ alloy, where $N^{X}_{d}$ is the number of $d$-electrons (X$=$A or B), it is convenient here to consider them as independent variables. Ordering of paramagnetic alloys occurs then when  $\overline{N}_{d}$ is comprised between 2.5 and 7.5. However, several alloys,  notably CoPt with a $\overline{N}_{d}$ of 8.3, order rather than phase separate for higher values of $\overline{N}_{d}$ \cite{Bieber91}.

Off-diagonal disorder defined as the difference between the band widths of the two constituents $\delta_{nd}=W^{A}_{d}-W^{B}_{d}$ could be relevant here because of its importance when mixing $3d$ and $5d$ elements.  Indeed, recently Los \textit{et al.} have shown that the variation of $V$ with $\overline{N}_{d}$ was extremely sensitive to $\delta_{nd}$~\cite{Los11}. Therefore for values of $\delta_{d}$ and $\delta_{nd}$ adapted to CoPt, the region of ordered alloys can be shown to be shifted towards higher values of $\overline{N}_{d}$, and can include CoPt. Thus, contrary to what was argued usually \cite{Ducastelle91} off-diagonal disorder can favor ordering instead of phase separation. However, the main reason for the ordering of CoPt and several other L1$_0$ structures seems to be the occurrence of ferromagnetism which  modifies the relevant number of $d$ electrons \cite{Bieber91}.

The role of magnetism can indeed be illustrated in a relatively simple manner in this scheme, by taking into account the strong ferromagnetic character of Co-Pt (Fig.~\ref{Figure4}c and d).  The majority spin up band is completely full (5 out of the total 8.3 $d$ electrons of the equiatomic CoPt are occupied) and does not participate to the cohesion of the system.  This results in considering a new effective average $\overline{N}_{d}$ equal to 3.3 $d$-electrons for a band normalized to 5 electrons or 6.6 electrons for a 10 electron band. This shift in the effective value of $\overline{N}_{d}$ due to magnetic considerations pushes the system fully into the area of $V>0$, which in turn explains the stability of ordered structures and the negative heats of formation present in this system. 
  
Actually it is certainly the presence of magnetic moments on Co that matters, more than the type (ferromagnetic, antiferromagnetic) of long range order which is stabilized at low temperature. This is clear when using the disordered local moment (DLM) picture where local moments interact through effective pair interactions \cite{Bieber91, Ruban04, Rahaman11}. 

To summarize, from the analysis of our \textit{ab initio} calculations within a simple tight-binding model,  we have shown that order in CoPt alloys,  in particular at the equiatomic concentration, is principally driven by magnetic effects. Off-diagonal disorder plays a role by shifting towards high values the range of band fillings where order is stabilized. Magnetism, more precisely the presence of local moments, strongly reinforces this tendency. This should equally apply to other similar alloys, antiferromagnetic coupling becoming competitive at lower band filling as in the case of FePt \cite{Lu10}.

\begin{acknowledgments}

Fruitful discussions with C. Barreteau, J. Los, and G. Tr\'eglia are gratefully acknowledged.
\end{acknowledgments}

\end{document}